\documentclass[preprint2,times,tighten]{aastex6}

\usepackage{graphicx,amssymb,aas_macros}
\usepackage{amsmath} 
\usepackage{amssymb} 
\usepackage{graphicx} 
\usepackage{indentfirst} 

\newcommand{\cmjjj}{\mbox{${\rm cm^{-3}}$}}
\newcommand{\etal}{et al.}

\newcommand{\kms}{\mbox{km\ s${^{-1}}$}}

\shorttitle{Detection of Extended HI in a $z=0.4$ Elliptical}
\shortauthors{Zahedy et al.}

\begin{document}


\title{{\it HST} Detection of Extended Neutral Hydrogen in a Massive Elliptical at $\lowercase{z}=0.4$}



\author{Fakhri S. Zahedy\altaffilmark{1,2}, 
Hsiao-Wen Chen\altaffilmark{1,3},
Michael Rauch\altaffilmark{2}, 
and Ann Zabludoff\altaffilmark{4}}

\altaffiltext{1}{Department of Astronomy \& Astrophysics, The University of Chicago, Chicago, IL 60637, USA} 
\altaffiltext{2}{The Observatories of the Carnegie Institution for Science, 813 Santa Barbara Street, Pasadena, CA 91101, USA}
\altaffiltext{3}{Kavli Institute for Cosmological Physics, The University of Chicago, Chicago, IL 60637, USA}
\altaffiltext{4}{Department of Astronomy, University of Arizona, Steward Observatory, Tucson, AZ 85721, USA}




\begin{abstract}

We report the first detection of extended neutral hydrogen (H\,I) gas in the interstellar
medium (ISM) of a massive elliptical galaxy beyond $z\sim0$. The observations 
utilize the doubly lensed images of QSO HE\,0047$-$1756 at $z_{\rm QSO}=1.676$ 
as absorption-line probes of the ISM in the massive ($M_{\rm star}\approx 10^{11}\, M_\odot$) 
elliptical lens at $z=0.408$, detecting gas at projected distances of $d=3.3$ and 4.6 kpc on 
opposite sides of the lens.  Using the Space Telescope Imaging Spectrograph 
(STIS), we obtain UV absorption spectra of the
lensed QSO and identify a prominent flux discontinuity and associated
absorption features matching the Lyman series transitions at $z=0.408$
in both sightlines.  The H\,I column density
is log $N\mathrm{(H\,I)}=19.6-19.7$ at both locations across the lens,
comparable to what is seen in 21 cm images of nearby ellipticals.  The
H\,I gas kinematics are well-matched with the kinematics of the Fe\,II
absorption complex revealed in ground-based echelle data, displaying a
large velocity shear of $\approx 360 \, \kms$ across the galaxy.  We
estimate an ISM Fe abundance of $0.3-0.4$ solar at both locations.
Including likely dust depletions increases the estimated Fe abundances
to solar or supersolar, similar to those of the hot ISM
and stars of nearby ellipticals.  Assuming 100\% covering fraction of
this Fe-enriched gas, we infer a total Fe mass of
$M_\mathrm{{cool}}({\rm Fe}) \sim (5-8)\times10^4 \ \,M_\odot$ in the
cool ISM of the massive elliptical lens, which is no more than 5\%
of the total Fe mass observed in the hot ISM.

\end{abstract}




\section[]{Introduction} 

A large fraction of massive quiescent galaxies are not gas-poor. 21~cm surveys
have revealed that $30-40$\% of of nearby early-type galaxies 
contain a large amount of neutral hydrogen (H\,I)
gas (e.g., Grossi \etal\ 2009; Oosterloo \etal\ 2010; Serra
\etal\ 2012; Young \etal\ 2014).  Beyond the local universe, QSO
absorption-line surveys of Mg\,II $\lambda\lambda$ 2796, 2803
absorption features near luminous red galaxies (LRGs)
reveal extended cool gas out to projected distances $d\gtrsim100$ kpc, 
with a mean covering fraction of $>15\%$ (e.g., Gauthier \etal\ 2009, 2010; Bowen \&
Chelouche 2011; Huang \etal\ 2016).  Because Mg\,II absorption traces
photoionized, cool $T\sim10^4\,$K gas (e.g., Bergeron \&
Stasi{\'n}ska 1986), the significant covering fraction of Mg\,II
absorbers indicates not only that cool gas is present in these
quiescent halos, but also that the gas has been enriched with heavy
elements.

\begin{figure*} 
\includegraphics[width=181mm]{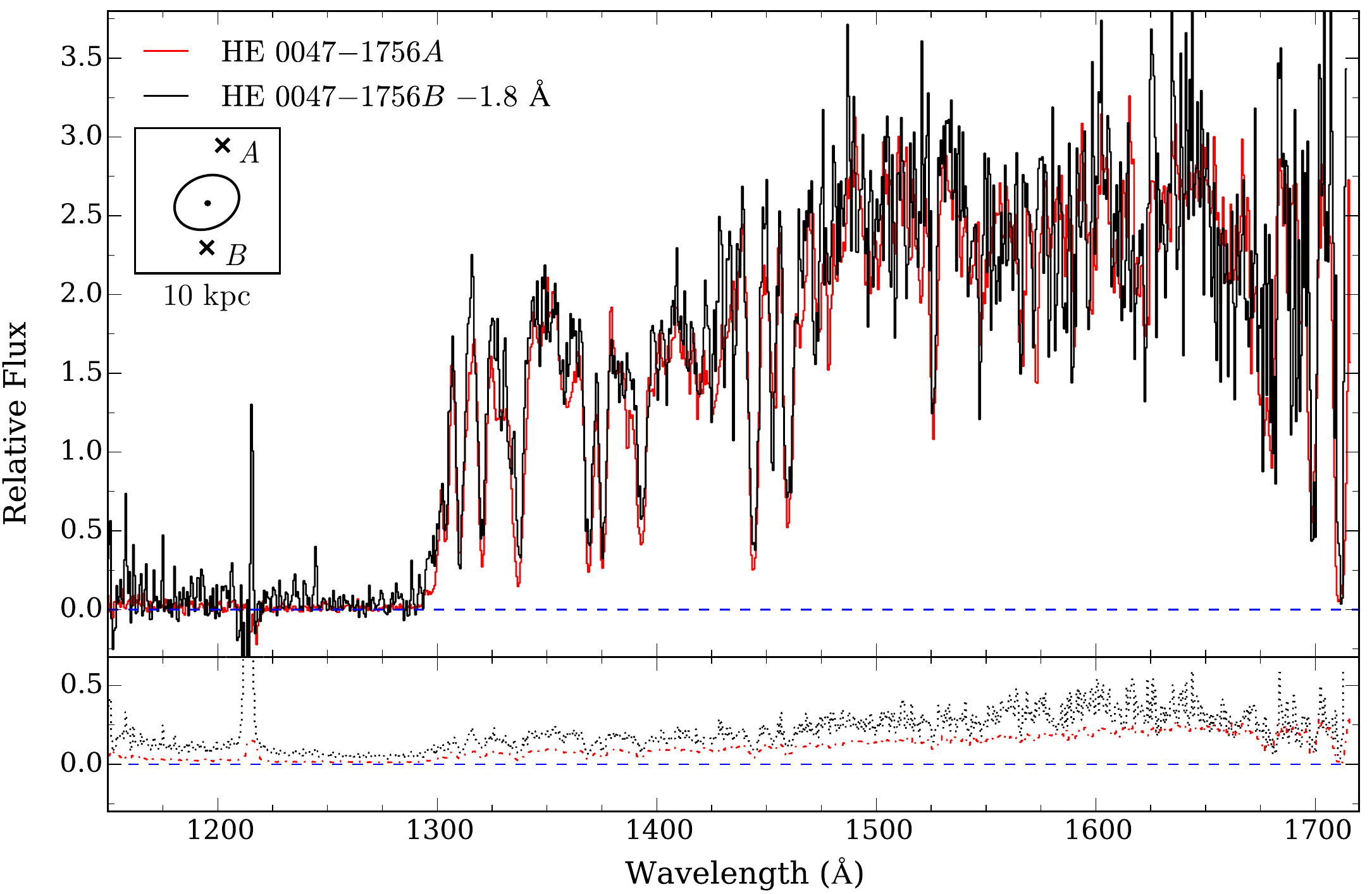}
\caption{{\it HST}/STIS FUV spectra of the doubly lensed
  QSO HE\,0047$-$1756.  The spectrum of HE\,0047$-$1756$B$ (red) has been shifted blueward 
  by $1.8\,$\AA\ to align the absorption features with those seen in
  HE\,0047$-$1756$A$ (black). 
  The corresponding 1-$\sigma$ error spectra are shown as 
  dotted lines in the bottom panel. The zero flux level is indicated with the blue 
  dashed line.  
  Inset: the configuration of the two lensed QSO sightlines (shown with crosses) in a $10 \times
  10$ kpc region around the lensing galaxy (shown with an ellipse).   The size and
  orientation of the ellipse correspond to the half-right radius and
  position angle of the semi-major axis of the lens galaxy as
  described in Z16.}
\label{Figure 1}
\end{figure*}

The finding that passive galaxies harbor a significant reservoir of
chemically enriched cool gas remains a central puzzle in
galaxy evolution, requiring some physical processes to prevent
the gas from cooling and eventually forming stars (e.g., McNamara \&
Nulsen 2007; Johansson \etal\ 2009; Conroy \etal\ 2015).  While 
feedback powered by central supermassive black
holes (SMBH) is frequently invoked as an explanation (e.g., McNamara
\& Nulsen 2007), observationally it has been difficult to directly
connect active galactic nuclei (AGN) activity to the quenching
of star formation.  Alternatively, the stars
themselves may be crucial in providing additional heating, which is
consistent with the observed anticorrelation between stellar surface
density and specific star formation rate in both low- and
high-redshift galaxies (e.g., Kauffmann \etal\ 2003, 2006; Franx
\etal\ 2008; Fang \etal\ 2013; Whitaker \etal\ 2017).  Specifically,
it has been suggested that old stellar populations can 
prevent continuing star-formation in passive
galaxies through energy injection from Type Ia supernovae (SNe Ia)
and winds from asymptotic giant branch (AGB) stars (see Conroy
\etal\ 2015 and references therein).

In a pilot study, Zahedy \etal\ (2016, hereafter Z16) employed
high-resolution absorption-line spectroscopy of lensed background QSOs
to examine the cool gas content in the inner halo ($d = 3-15$ kpc) of
three massive quiescent lensing galaxies at $z=0.4-0.7$.  While the
gas content varied significantly among lenses
and different sightlines of the same lens, a supersolar
$\mathrm{Fe/Mg}$ relative abundance pattern,
$[\mathrm{Fe}/\mathrm{Mg}]\gtrsim0.1$, was found in {\it every}
sightline where cool gas was detected ($\sim40$\% of all sightlines).
Such high $\mathrm{[Fe/Mg]}$ ratios indicate a significant
contribution ($\gtrsim 20$\%) from SNe Ia to the chemical enrichment
history of the gas, exceeding what have been observed in mature
stellar populations such as the solar neighborhood (e.g., Tsujimoto
\etal\ 1995). Z16 suggest that these absorbers reside
in the interstellar medium (ISM), where the gas is likely to have been
enriched and heated by SNe Ia ejecta.  However, to facilitate a direct
comparison with different chemical enrichment models, it is necessary
to probe the bulk of the cool ISM, neutral hydrogen (H\,I), and
constrain its column density, $N({\rm H\,I})$.

We have obtained UV absorption spectra of the doubly lensed QSO
HE\,0047$-$1756 at $z_{\rm QSO}=1.676$, using the Space Telescope
Imaging Spectrograph (STIS) on board the {\it Hubble Space Telescope}
({\it HST}).  The lens of HE\,0047$-$1756 at $z=0.408$ is a massive elliptical galaxy
with a total stellar mass of $M_*\approx 10^{11}\,
M_\odot$ and no ongoing star formation (Z16).  Ultra-strong Mg\,II absorbers
of rest-frame absorption equivalent width
$W_r(2796)>3.6\,\mathrm{\AA}$ are found along both lensed QSO
sightlines probing opposite sides of this LRG lens at $d = 4.6$
kpc (1.8 half-light radii, $r_e$) and $d = 3.3$ kpc (1.3 $r_e$).
The STIS spectra allow us to measure the total integrated $N\mathrm{(H\,I)}$ at two
locations separated by $\approx 8$ kpc in this massive quiescent
galaxy, thereby constraining the mean gas metallicity.  In
this Letter, we report a large column density of
extended neutral hydrogen gas detected in the inner ISM of the HE\,0047$-$1756
lens, the first such discovery in a quiescent galaxy beyond $z\sim0$.
Throughout this Letter, we adopt a flat cosmology of $\Omega_{\rm
  M}=0.3$ and $\Omega_\Lambda = 0.7$, with a Hubble constant of $H_0 =
70 \ {\rm km} \ {\rm s}^{-1}\ {\rm Mpc}^{-1}$.

\section[]{Observations and Data Reduction} 

Long-slit FUV spectroscopy of the doubly lensed QSO HE\,0047$-$1756
was obtained on 2016 December 5 using the FUV-MAMA detector in
{\it HST}/STIS with the $52''$$\times$ $0''.2$ long slit and the
low-resolution G140L grating (PID: 14751; PI: Zahedy).  This
configuration provides a dispersion of $\approx 0.6$ \AA\ per pixel
with a corresponding spectral resolution of $\approx 270 \,\kms$.
Because the two lensed images of HE\,0047$-$1756 are separated by
merely $1.4''$, the long-slit spectroscopic mode of STIS enabled
observation of both QSO images in a single setup, which was achieved by orienting the slit
 at a position angle of
$-9.51^\circ$.  At the redshift of the lensing galaxy, the broad
wavelength coverage of the G140L grating ($1150-1720$ \AA) includes
the entire H\,I Lyman series.
The total integration time of the observations was 7689 s, which was
divided into six individual exposures of roughly equal durations.

\begin{figure*} 
\hspace{-0.5em}
\includegraphics[width=181mm]{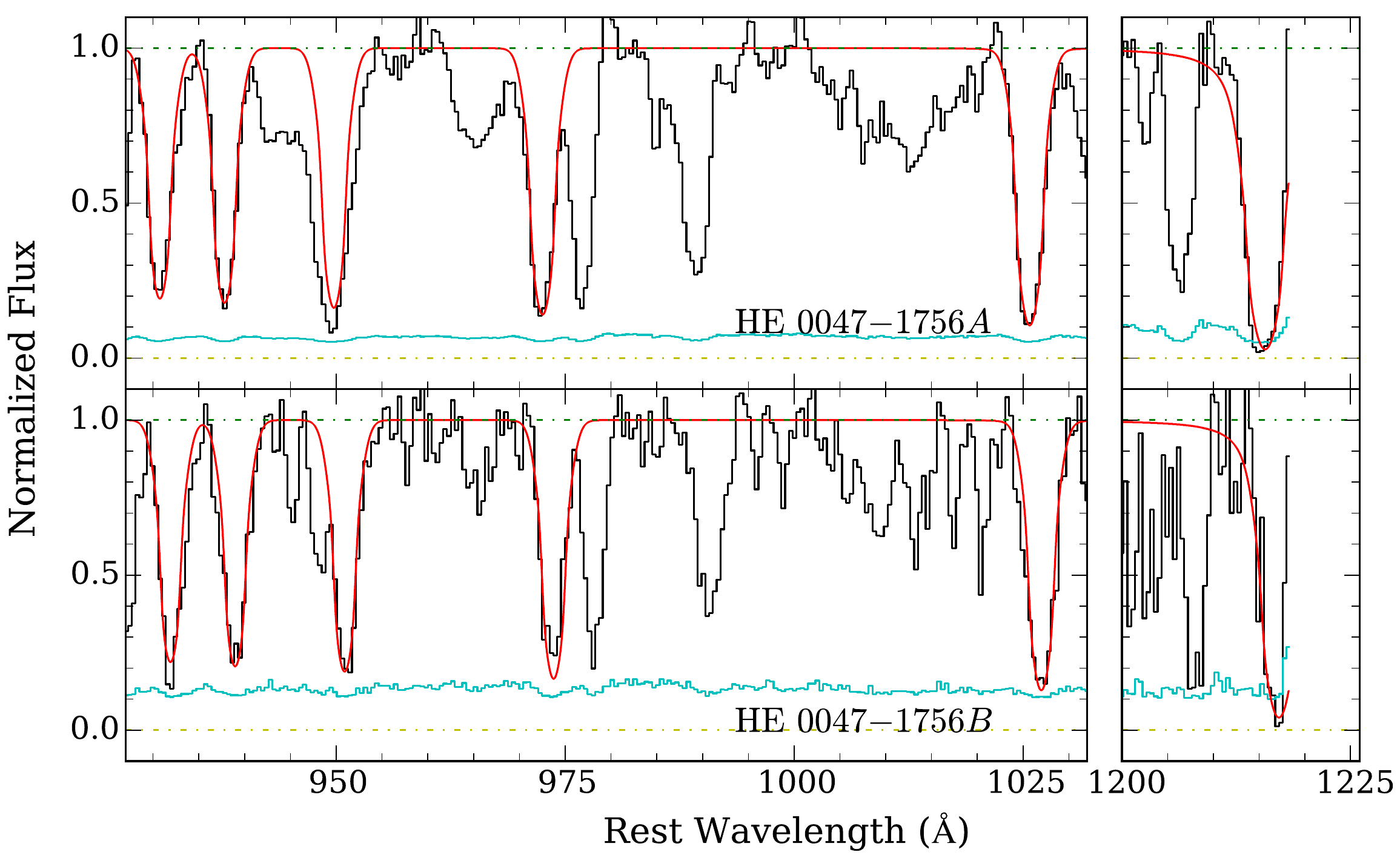}
\caption{Continuum-normalized absorption profile of H\,I Lyman series
  transitions along the doubly lensed QSO sightlines HE\,0047$-$1756$A$
  (top) at $d$= 4.6 kpc (or 1.8 $r_e$) and HE\,0047$-$1756$B$ (bottom)
  at $d$= 3.3 kpc (or 1.3 $r_e$) from the HE\,0047$-$1756 lens
  galaxy. The x-axis corresponds to the rest-frame wavelength of
  the galaxy at $z=0.408$. The 1$\sigma$ error spectra are plotted in
  cyan histograms. The best-fit Voigt-profile models for the H\,I
  Lyman series, convolved with the STIS LSF, are included in red, with log
  $N\mathrm{(H\,I)}=19.7^{+0.1}_{-0.1}$ for the absorber along
  HE\,0047$-$1756$A$ and log $N\mathrm{(H\,I)}=19.6^{+0.2}_{-0.3}$
  for the absorber along HE\,0047$-$1756$B$.   Note that
  the Ly$\delta \,\lambda949$ line is contaminated by
  C\,II$\,\lambda1334$ absorption from the Milky Way ISM.  }
\label{Figure 2}
\end{figure*}

Individual spectral images were initially processed using the
automatic STIS calibration pipeline, \textsc{Calstis} (Hodge
\etal\ 1998), which performed dark and bias
subtractions, flat-fielding, and wavelength calibration, and produced
a series of data products including a two-dimensional rectified
spectral image for each science exposure.  The pipeline-calibrated
data products were retrieved from the {\it HST}
archive for further processing.  We developed a custom spectral
extraction software to optimize signal-to-noise ratio (S/N) in the
resulting spectra, especially that of the fainter lensed QSO image
HE\,0047$-$1756$B$.  The custom software first constructed
a model spatial profile along the cross dispersion direction for each
QSO image by averaging the two-dimensional rectified spectral image
along the wavelength direction.  Regions affected by strong sky
emission lines were excluded from this exercise.  Next, the software
extracted individual QSO spectra by applying the mean spatial profile
of each QSO image as optimal weights for the spatially resolved
spectral signal at each wavelength.  Finally, the optimally extracted
one-dimensional spectra from individual exposures were coadded to
form a single combined spectrum per QSO image, which is characterized
by a mean S/N $\approx 20-25$ per resolution element for
HE\,0047$-$1756$A$, and S/N $\approx10-13$ per resolution element for
HE\,0047$-$1756$B$, over the full wavelength range ($1150-1720$ \AA).

\section[]{Large Interstellar $N\mathrm{(HI)}$ in the Massive Quiescent Lens}

The final coadded FUV spectra of the two lensed QSO images are
presented in Figure 1.  The absorption features along
sightline $B$ (black) are redshifted by $\approx1.8\,$\AA\ relative to those
along sightline $A$ (red), consistent with the kinematic offset of $\approx
360 \, \kms$ seen in Mg\,II, Fe\,II, and Mg\,I.  Strong absorption features are detected along both
sightlines, which can be attributed to the H\,I Lyman series as well
as various low- to high-ionization metal transitions at the redshift
of the lensing galaxy. 

For both sightlines, the FUV spectra exhibit a lack of continuum flux 
blueward of the Lyman limit (at $\approx 1280$ \AA\ in observed wavelength) 
expected for the lens galaxy.  The mean flux averaged over a finite spectral window below the 
Lyman limit is consistent with zero to within the 1$\sigma$ associated uncertainty, indicating
that the absorbing gas at $d\sim r_e$ from the lens galaxy of
HE\,0047$-$1756 is optically thick to hydrogen-ionizing photons. 
We estimate a 2$\sigma$ upper limit to the
Lyman-continuum flux using the mean value of the error spectrum
between 870 and 900 \AA\ in the rest frame of the lens galaxy, which we 
then convert to a 2$\sigma$ lower limit to the continuum opacity by
calculating its ratio to the mean continuum level redward of the Lyman
limit at rest-frame $955-965$ \AA. Given that the H\,I
photoionization cross-section varies as $\sigma(\lambda)=
6.30\times10^{-18} \,(911.8\, \mathrm{\AA}/\lambda)^{-3}
\,\mathrm{cm^2}$ (Osterbrock \& Ferland 2006), we estimate 2$\sigma$
lower limits of log $N\mathrm{(HI)}>17.8$ for HE\,0047$-$1756$A$ and
log $N\mathrm{(HI)}>17.6$ for HE\,0047$-$1756$B$ at the lens redshift.

Including the entire Lyman series lines allows us to better constrain
$N\mathrm{(HI)}$.  We perform a Voigt-profile analysis using a custom
program previously developed by and described in Z16.  For each
sightline, the combined spectrum is first continuum normalized using a
low-order polynomial function.  Then, a simultaneous Voigt-profile fit
is performed on the following Lyman series lines: Ly$\alpha
\,\lambda1215$, Ly$\beta \,\lambda1025$, Ly$\gamma \,\lambda972$,
Ly$\epsilon\,\lambda937$, and Ly$\zeta\,\lambda930$.  The Ly$\delta
\,\lambda949$ line is excluded from the fit because it is blended with
the Galactic C\,II$\,\lambda1334$ absorption.  Higher-order Lyman
series lines (Ly-7 and onward) are also excluded due to contaminations
by the [O\,I] $\lambda\lambda$ 1302,1306 airglow emission-line
doublet.

The Voigt-profile analysis results in a best-fit log
$N\mathrm{(H\,I)}=19.7\pm0.1$ for the absorber along
HE\,0047$-$1756$A$, and log $N\mathrm{(H\,I)}=19.6^{+0.2}_{-0.3}$
for the absorber along HE\,0047$-$1756$B$.  The quoted uncertainties
represent the estimated 95\% confidence level of the best-fit log
$N\mathrm{(H\,I)}$, marginalized over the Doppler parameter $b$. 
The best-fit Voigt profiles of the Lyman series 
lines are presented in Figure 2 (in red) along with the data (in black).  A lower N(HI) value fails to
reproduce the damping wing of the Ly$\alpha$ line, demonstrating that
strong constraints can be obtained for $N\mathrm{(H\,I)}$ based on the
low-resolution STIS spectra.  We also perform a curve-of-growth (COG)
analysis for estimating $N\mathrm{(H\,I)}$ along each sightline and
find consistent results with the Voigt-profile analysis, with log $N\mathrm{(H\,I)}=19.6\pm0.2$
for A and log $N\mathrm{(H\,I)}=19.5^{+0.2}_{-0.4}$ for B, and 
an effective $b$ value of $135\pm 10 \, \kms$ for A and $125\pm15 \, \kms$ for B.
The large effective $b$ values are understood to be driven by gas kinematics 
rather than the actual thermal width of the gas. 
To quantify the uncertainties in $N\mathrm{(H\,I)}$ and $b$ as a result of uncertainties in the STIS
line-spread function (LSF), we 
repeat the Voigt-profile analysis by varying the width of the STIS LSF, and find
that a 10\% change in the LSF results in less than a 5\% change in the best-fit $b$
and less than a 10\% change in $N\mathrm{(H\,I)}$.

\section[]{Discussion and Implications}

Our analysis of the STIS spectra of the doubly lensed QSO
HE\,0047$-$1756 has yielded, for the first time, constraints on the
neutral gas content at two locations separated by $\approx 8$ kpc in
a massive quiescent galaxy beyond $z\sim0$. The range of observed
$N\mathrm{(H\,I)} $ along both lensed sightlines, log
$N\mathrm{(H\,I)}=19.6-19.7$, is comparable to what is commonly seen in 21cm
maps of the ISM in nearby early-type galaxies (log
$N\mathrm{(H\,I)}\approx 19.3-20.3$ e.g., Oosterloo \etal\ 2010;
Serra \etal\ 2012), but is high among previous
single-sightline surveys of more distant passive galaxies
(e.g., Thom \etal\ 2012; Prochaska \etal\ 2017).
Detections of high-$N\mathrm{(H\,I)}$ absorbers at two different
locations suggest the presence of extended interstellar H\,I.  Here, we
discuss the implications of this finding.

\subsection[]{Chemical enrichment level in the cool ISM of the lens}

While the detailed gas kinematics are not resolved in the
low-resolution spectra, the line centroids can be determined to high
accuracy.  The H\,I gas kinematics are found to match well with the
kinematics of the Fe\,II absorption complex revealed in ground-based
echelle data.  Both H\,I and metal absorption lines display a large
velocity shear of $v_B -v_A= +360 \, \kms$ over a projected
distance of $\approx 8$ kpc across the quiescent galaxy, flanking the
systemic redshift of the lens.

We estimate a mean gas metallicity averaged across all components
 by combining the newly available
$N\mathrm{(H\,I)}$ and previously published total Fe\,II column
densities from Zahedy \etal\ (2017). The Fe abundance of the gas,
$\mathrm{(Fe/H)}$, is related to the observed column density ratio of
Fe\,II and H\,I corrected for the expected ionization fractions,
following $\log\,({\rm Fe}/{\rm H}) =
\log\,N\mathrm{(Fe\,II)}/N\mathrm{(H\,I)} -
\log\,\mathrm{(\mathit{f}_{Fe^+}/\mathit{f}_{H^0})}$.  In the absence
of a large {\it relative} ionization fraction correction, with
$\log\,\mathrm{(\mathit{f}_{Fe^+}/\mathit{f}_{H^0})}\sim 0$, the
observed column density ratio translates directly into the Fe
abundance, $\log\,N\mathrm{(Fe\,II)}/N\mathrm{(H\,I)}\approx\log\,({\rm
  Fe}/{\rm H})$.

We perform a series of photoionization calculations using
\textsc{Cloudy} v.13.03 (Ferland \etal\ 2013) to estimate the
necessary ionization fraction corrections for a wide range of gas
densities and metallicities in a photoionized gas of temperature
$T=10^4$ K and $\log\,N\mathrm{(H\,I)}=19.6-19.7$.  We assume a
plane-parallel slab, illuminated on both sides with the updated Haardt
\& Madau (2001) ionizing background radiation field (HM05 in
\textsc{Cloudy}) at $z = 0.4$.  Adopting a maximum cloud size leads to
a lower limit on the allowed gas density for a given
$N\mathrm{(H\,I)}$.  We adopt a maximum cloud size of 1 kpc,
consistent with observations of Galactic high-velocity clouds (e.g.,
Putman \etal\ 2012), and find $n_{\rm H}\gtrsim 10^{-1.6}$ cm$^{-3}$
for the absorbers along the two lensed QSO sightlines.  Over this
allowed gas density range, the calculations show that the gas is
mostly neutral with $f_{\rm H^0}>0.6$ and the relative ionization
fraction correction between Fe$^{+}$ and H$^0$ is
$\mathrm{(\mathit{f}_{Fe^+}/\mathit{f}_{H^0})\approx 1}$.  Therefore,
to a good approximation, the gas metallicity can be estimated directly
from $\log\,N\mathrm{(Fe\,II)}/N\mathrm{(H\,I)}$, following
$\mathrm{[Fe/H]\approx log\,\mathit{N}(Fe\,II)/\mathit{N}(H\,I) -
  log\,(Fe/H)_\odot}$\footnote{If the intrinsic $N\mathrm{(H\,I)}$ of the Fe\,II absorbing gas is lower than observed (namely, only part of
the observed $N\mathrm{(H\,I)}$ is cospatial with the Fe\,II gas), then
the gas
metallicity would be higher still.}. 

The total Fe\,II column density is log
$N\mathrm{(Fe\,II)}=14.61\pm0.02$ for the absorber along
HE\,0047$-$1756$A$ and log $N\mathrm{(Fe\,II)}=14.70\pm0.08$ for the
absorber along HE\,0047$-$1756$B$ (Z16; Zahedy \etal\ 2017).  Adopting
the solar chemical abundance pattern from Asplund \etal\ (2009), where
log\,$\mathrm{(Fe/H)_\odot=-4.50\pm0.04}$, we find that the estimated
Fe abundance is $\mathrm{[Fe/H]}=-0.6\pm0.1$ for the absorber along
HE\,0047$-$1756$A$, and $\mathrm{[Fe/H]}=-0.4\pm0.3$ for the absorber
along HE\,0047$-$1756$B$.  

\subsection[]{Effects of dust depletion}

Dust depletion, when unaccounted for, can lead to
underestimated chemical abundances, particularly in high-metallicity
gas.  For the HE\,0047$-1756$ lens, Ca\,II $\lambda\lambda$ 3934,3969
absorption has also been observed along both sightlines (Z16).  It has
been shown that the strength of Ca\,II absorption correlates with both
reddening (Murga \etal\ 2015) and the $\mathrm{[Cr/Zn]}$ relative
abundance (e.g., Zych \etal\ 2009), two common diagnostics of the dust
content.  Z16 previously reported $W_r(3934)=0.3\,\mathrm{\AA}$ and
$W_r(3934)=0.1\,\mathrm{\AA}$ for the absorption systems along
sightlines $A$ and $B$, respectively. For moderately strong Ca\,II
absorbers of $W_r(3934)=0.1-0.5\,\mathrm{\AA}$, the range of
observed $\mathrm{[Cr/Zn]}$ is $\mathrm{[Cr/Zn]\sim-0.4\pm 0.2 }$
(e.g., Zych \etal\ 2009 and references therein). The inferred
$\mathrm{[Cr/Zn]}$ level indicates a modest amount of dust depletion similar to
that seen in the Galactic halo, where the expected amount of Fe
depletion is $0.5-0.6$ dex (e.g., Savage \& Sembach 1996; De Cia
\etal\ 2016). For both sightlines, applying
the expected dust depletion leads to still higher gas metallicities,
reaching solar or supersolar values, $\mathrm{[Fe/H]}\gtrsim0$, and
comparable to the typical mean metallicities in the hot ISM of nearby
ellipticals, $\mathrm{[Fe/H]}\gtrsim0-0.3$ (e.g., Humphrey \&
Buote 2006; Loewenstein \& Davis 2010, 2012).

\subsection[]{Fe Mass Budget}

In Z16, we reported a supersolar $\mathrm{[Fe/Mg]}$ abundance pattern
in the lensing galaxy and suggested that the gas has been enriched by
the Fe-rich ejecta from SNe Ia.  
Now that we have estimated the gas metallicity, we can estimate the total Fe mass
observed in the cool ISM and compare it with the
expected contribution from SNe Ia.

The Fe mass in the cool ISM of the lens is estimated based on the
observed $N\mathrm{(H\,I)}$, the inferred solar gas metallicity, and the ionization state of the gas
from the \textsc{Cloudy} models in \S\ 4.1.  We find a total Fe mass
within $d=5$ kpc ($\approx 2\,r_e$; the region probed by the doubly
lensed QSO) of $M_\mathrm{{cool}}({\rm Fe})\sim (5-8)\times10^4 \,
(f_\mathrm{cov}) \,M_\odot$ in the cool ISM of the lens, where $f_\mathrm{cov}$
is the cool gas covering fraction. 

Next, we estimate the expected SNe Ia contribution based on the
observed rate and expected yields of SNe Ia.  The observed radial
distribution of SNe Ia in elliptical galaxies follows the stellar
light distribution (F{\"o}rster \& Schawinski 2008). The
mean SNe Ia rate per unit stellar mass is found to be 0.044 SNe per
century per $10^{10}\,M_\odot$ in nearby ellipticals (Mannucci
\etal\ 2005).  Combining these findings, we estimate that the integrated SNe Ia 
rate within $d\sim5$ kpc of the lens is $\sim0.3$ per century. 
Given a minimum stellar population age of $\approx 1$ Gyr for this
galaxy (Z16) and assuming a constant SNe Ia rate, a total of $3\times 10^6$ SNe Ia 
should have occurred in this volume. Because SN Ia is expected to 
produce $\approx 0.6-0.7\,M_\odot$ in Fe (e.g., Iwamoto \etal\ 1999),
the estimated total number of SNe Ia leads to an expected total Fe mass of
$M_\mathrm{Fe_{Ia}} \sim 2\times10^6 \,M_\odot$ over the 1 Gyr time
interval.  Note that if the SNe Ia rate were higher in the past, then
this estimate should be considered as a lower limit.  Comparing
$M_\mathrm{{cool}}({\rm Fe})$ to $M_\mathrm{Fe_{Ia}}$, we conclude
that the Fe uncovered in the absorbers accounts for at most $\sim 5$\% of all 
Fe produced by SNe Ia over the lifetime of a massive
elliptical galaxy.  As a result,  the majority of Fe must reside in a hot ISM phase.

Although no direct observations of the hot ISM are available for this
quiescent galaxy, we could use nearby massive ellipticals as a
reference.  X-ray observations have shown that their hot ISM has
 a typical mean $\mathrm{[Fe/H]}\sim0$ and
a mean central density of $n_\mathrm{H}\sim 0.1\, \cmjjj$ (e.g.,
Mathews \& Brighenti 2003).  Adopting these numbers, we infer a total
hot-phase Fe mass within $2\,r_e$ of $M_\mathrm{{hot}}({\rm
 Fe}) \sim 2.5\times10^6 \,M_\odot$, in good agreement with the 
 expected contribution from SNe Ia.  Therefore,
the implication that the majority of Fe produced in SNe Ia resides in
the hot phase is at least consistent with the empirical understanding
of nearby massive ellipticals.

\section[]{Conclusions}

Using{\it HST}/STIS UV spectra of the doubly lensed images of QSO HE\,0047$-$1756, 
we have discovered a high column density of neutral hydrogen gas at $3<d<5$ kpc (or
$r_e<d<2\,r_e$) from a massive ($M_{\rm star}\approx 10^{11}\, M_\odot$)
elliptical lens galaxy at $z=0.408$. The new STIS spectra not only allow direct measurements of the
ISM $N({\rm HI})$ and metallicity, but also constrain the Fe mass budget in this distant elliptical.  
We find an ISM $N\mathrm{(H\,I)}$ of log $N\mathrm{(H\,I)}=19.6-19.7$
at two locations separated by $\approx 8$
kpc on opposite sides of the galaxy.  This large $N\mathrm{(H\,I)}$ is
comparable to the range of $N\mathrm{(H\,I)}$ in nearby
ellipticals.  The Fe abundance is $0.3-0.4$ solar along both sightlines, which
 increases to solar or supersolar after accounting for likely
dust depletions. Together, the observed $N({\rm HI})$ and Fe abundance constrain the Fe mass budget.
For a 100\% gas covering fraction, 
we infer a total Fe mass of $M_\mathrm{{cool}}({\rm Fe}) \sim
(5-8)\times10^4 \ \,M_\odot$ in the cool ISM of the lens.  
Including previous findings that SNe Ia contribute
significantly to the chemical enrichment in massive quiescent halos (e.g., Mernier \etal\ 2017),
we find that the majority ($\approx 95$\%) of Fe produced by SNe Ia
resides in the hot phase.  Future studies probing
warm ($T\sim10^{5-6}$ K) gas in the halos of luminous red galaxies 
using high-ionization metal absorption (e.g., O\,VI, likely tracing cooling gas) will provide
new insights into SNe Ia heating and assess its
relative importance to other forms of late-time feedback in the halos of massive
quiescent galaxies.

\acknowledgements

We thank Annalisa de Cia, Sean Johnson, and John Mulchaey  for helpful
discussions. F.S.Z. and H.W.C. acknowledge partial support from
HST-GO-14751.01A. F.S.Z. acknowledges support from the Brinson Foundation
predoctoral fellowship. A.I.Z. acknowledges support from NSF grant
AST-1211874. This work is based on data gathered with the NASA/ESA
{\it Hubble Space Telescope} operated by the Space Telescope Science
Institute and the Association of Universities for Research in
Astronomy, Inc., under NASA contract NAS 5-26555.



\end{document}